\documentclass[reprint, floatfix, superscriptaddress, prx, aps]{revtex4-2}

\usepackage[table]{xcolor}
\usepackage{stmaryrd}
\usepackage{physics}
\usepackage{graphicx}
\usepackage{svg}
\usepackage{listings}
\usepackage[breaklinks]{hyperref}
\usepackage[capitalize]{cleveref}
\usepackage{nicefrac}
\usepackage{placeins}
\usepackage{array}
\usepackage{multirow}
\usepackage{soul}
\definecolor{tikzwhite}{RGB}{255,255,255}
\definecolor{tikzblack}{RGB}{0,0,0}
\definecolor{tikzgray}{RGB}{128,128,128}
\definecolor{tikzyellow}{RGB}{255,255,0}
\definecolor{tikzmagenta}{RGB}{255,0,255}
\definecolor{tikzred}{RGB}{255,0,0}
\definecolor{tikzcyan}{RGB}{0,255,255}
\definecolor{tikzgreen}{RGB}{0,255,0}
\newcommand{\logicallabel}[1]{{\fontsize{0.8cm}{1cm} #1}}

\newcommand{\nkd}[3]{\left \llbracket #1,\, #2,\, #3 \right \rrbracket}
\newcommand{\ccz}{\textsc{ccz}}
\newcommand{\cnot}{\textsc{cnot}}
\newcommand{\swap}{\textsc{swap}}

\newcommand{\ct}{^{\dagger}}
\newcommand{\tgate}{\textsc{t}}
\newcommand{\SICC}{\nkd{8}{3}{2}}
\usepackage{tikz}
\usetikzlibrary{decorations.pathreplacing,decorations.pathmorphing,arrows.meta}

\lstdefinelanguage{OpenQASM} {
  morekeywords={qreg,creg,include, gate, opaque,U,CX,measure, reset,if,barrier,
    u3,u2,u1,cx,id,x,y,z,h,s,sdg,t,tdg,rx,ry,rz,cz,cy,ch,ccx,crz,cul,cu3},
  sensitive=true,
  morecomment=[l]{//},
  morestring=[b]"
}

\begin{document}
	\title{Fault-Tolerant One-Bit Addition with the Smallest Interesting Colour Code}
	\author{Yang Wang}
	\affiliation{QuTech, Delft University of Technology, PO Box 5046, 2600 GA Delft, The Netherlands}
	\affiliation{3. Physikalisches Institut, ZAQuant, University of Stuttgart, Allmandring 13, 70569 Stuttgart, Germany}	
	\author{Selwyn Simsek}
	\affiliation{Quantinuum, Terrington House, 13–15 Hills Road, Cambridge, CB2 1NL, UK}
	\author{Thomas M. Gatterman}
	\author{Justin A. Gerber}
	\author{Kevin Gilmore}
	\author{Dan Gresh}
	\author{Nathan Hewitt}
	\author{Chandler V. Horst}
	\author{Mitchell Matheny}
	\author{Tanner Mengle}
	\author{Brian Neyenhuis}
	\affiliation{Quantinuum, 303 South Technology Ct., Broomfield, CO 80021, USA}
	\author{Ben Criger}
	\email{ben.criger@quantinuum.com}
	\affiliation{Quantinuum, Terrington House, 13–15 Hills Road, Cambridge, CB2 1NL, UK}
	\date{\today}
	\begin{abstract}
Fault-tolerant operations based on stabilizer codes are the state of the art in suppressing error rates in quantum computations.
Most such codes do not permit a straightforward implementation of non-Clifford logical operations, which are necessary to define a universal gate set.
As a result, implementations of these operations must either use error-correcting codes with more complicated error correction procedures or gate teleportation and magic states, which are prepared at the logical level, increasing overhead to a degree that precludes near-term implementation.
In this work, we implement a small quantum algorithm, one-qubit addition,
fault-tolerantly on the Quantinuum H1-1 quantum computer, using the $\SICC$ colour code.
By removing unnecessary error-correction circuits and using low-overhead techniques for fault-tolerant preparation and measurement, we reduce the number of error-prone two-qubit gates and measurements to 36.
We observe arithmetic errors with a rate of $\sim 1.1 \times 10^{-3}$ for the fault-tolerant circuit and $\sim 9.5 \times 10^{-3}$ for the unencoded circuit.
\end{abstract}

\maketitle
\section{Introduction} \label{sec:intro}
Quantum computers have a large and growing number of potential applications \cite{montanaroQuantumAlgorithmsOverview2016}, and quantum computers of increasing size are being constructed \cite{pinoDemonstrationTrappedionQuantum2021,IBMUnveilsBreakthrough}.
Owing to the effects of noise and physical imperfections, these devices continue to have large physical error rates, on the order of $2 \times 10^{-3}$ \cite{QuantinuumSystemModel2022} for a typical two-qubit entangling gate or measurement, preventing the direct implementation of large-scale algorithms with physical qubits.
Therefore, it is necessary to carry out a quantum computation fault-tolerantly to lessen the effective error rate.

One straighforward way to construct a fault-tolerant circuit is to select a quantum error-correcting code together with a corresponding set of fault-tolerant operations, and replace the individual operations of a given non-fault-tolerant circuit with their fault-tolerant counterparts \cite{shorFaulttolerantQuantumComputation1996}.
Error-correction gadgets designed for the code in question are then inserted between these fault-tolerant operations to prevent the effect of errors building up over time.
Indeed, proofs of the threshold theorem often provide an explicit construction of such a procedure \cite{aliferisQuantumAccuracyThreshold2005}.
However, this procedure often involves a large overhead in terms of both gate
count and number of qubits required, which limits the suitability of such
circuits to be implemented on experimental hardware. One consequence of this overhead is that although the ingredients of fault tolerant computation including memory gadgets
\cite{ryan-andersonRealizationRealTimeFaultTolerant2021}, entangling gates
\cite{ryan-andersonImplementingFaulttolerantEntangling2022} and preparation of
magic states \cite{postlerDemonstrationFaulttolerantUniversal2022} have been
demonstrated individually, they have generally not yet been implemented at the
same time.

Reducing the size of a fault-tolerant circuit (in terms of gate and qubit count)
can make implementation easier. Smaller circuits are also preferable as there is reason to expect that they result in lower logical error rates.
Consider two fault-tolerant circuits $C_s$ and $C_l$ that act identically on all
inputs, where the circuit $C_s$ is shorter than the circuit $C_l$, and both
circuits can detect or correct $t-1$ faults. $C_l$ contains more locations at which faults may occur and, therefore, has a higher likelihood of experiencing at least $t$ faults, which may be undetectable/uncorrectable and contribute to the logical error rate, provided all other factors are equal.
This is another reason why smaller fault-tolerant circuits are to be preferred
over larger ones.

Much previous work employs a conventional approach to implementing fault-tolerant algorithms.
One starts with a code in which it is straightforward to implement Clifford gates directly.
For instance, this can be done with surface code patches via lattice surgery
\cite{litinskiLatticeSurgeryTwist2018}, and we note that an instance of Grover's algorithm not containing non-Clifford gates has already been implemented fault-tolerantly on two qubits \cite{pokharelBetterthanclassicalGroverSearch2022} using the $\nkd{4}{2}{2}$ code.
Then, to obtain a universal gate set, a single non-Clifford gate such as the $\tgate$ gate is implemented by gate teleportation \cite{gottesmanQuantumTeleportationUniversal1999}, which may require magic state distillation \cite{bravyiUniversalQuantumComputation2005}.

A recent proposal \cite{landahlPersonalCommunication2022} turns this scenario on its head. One may instead start with codes which possess transversal (and hence fault-tolerant) non-Clifford gates.
The Clifford gates are then implemented by gate teleportation with Pauli eigenstates as the ancillary inputs, requiring no expensive distillation to prepare.

In this paper, we realise a small instance of this proposal.
We implement a one-qubit addition circuit using the $\SICC$ colour code, which allows a transversal non-Clifford $\ccz$ gate.
Although this algorithm is simple, it computes the answer to a mathematical
problem and contains both Clifford and non-Clifford gates.

The one-bit adder circuit is also a simple application of the Toffoli gate, which is a universal gate for reversible classical computing.
It can be used to construct oracles for Grover’s algorithm \cite{groverFastQuantumMechanical1996,pokharelBetterthanclassicalGroverSearch2022} as well as the modular exponentiation circuits employed within Shor’s algorithm.
As a result, the fault-tolerant realisation of a one-bit adder circuit has
practical implications for the realisation of more substantial quantum
algorithms, as well as pushing forward the state-of-the-art in the realisation
of fault tolerant algorithms on contemporary quantum computers. 
\section{One-Bit Addition}
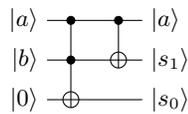
\begin{figure}
  \begin{tikzpicture}[scale=1.000000,x=1pt,y=1pt]
\filldraw[color=white] (0.000000, -7.500000) rectangle (36.000000, 37.500000);
\draw[color=black] (0.000000,30.000000) -- (36.000000,30.000000);
\draw[color=black] (0.000000,30.000000) node[left] {$|a\rangle$};
\draw[color=black] (0.000000,15.000000) -- (36.000000,15.000000);
\draw[color=black] (0.000000,15.000000) node[left] {$|b\rangle$};
\draw[color=black] (0.000000,0.000000) -- (36.000000,0.000000);
\draw[color=black] (0.000000,0.000000) node[left] {$|0\rangle$};
\draw (9.000000,30.000000) -- (9.000000,0.000000);
\filldraw (9.000000, 30.000000) circle(1.500000pt);
\filldraw (9.000000, 15.000000) circle(1.500000pt);
\begin{scope}
\draw[fill=white] (9.000000, 0.000000) circle(3.000000pt);
\clip (9.000000, 0.000000) circle(3.000000pt);
\draw (6.000000, 0.000000) -- (12.000000, 0.000000);
\draw (9.000000, -3.000000) -- (9.000000, 3.000000);
\end{scope}
\draw (27.000000,30.000000) -- (27.000000,15.000000);
\filldraw (27.000000, 30.000000) circle(1.500000pt);
\begin{scope}
\draw[fill=white] (27.000000, 15.000000) circle(3.000000pt);
\clip (27.000000, 15.000000) circle(3.000000pt);
\draw (24.000000, 15.000000) -- (30.000000, 15.000000);
\draw (27.000000, 12.000000) -- (27.000000, 18.000000);
\end{scope}
\draw[color=black] (36.000000,30.000000) node[right] {$|a\rangle$};
\draw[color=black] (36.000000,15.000000) node[right] {$|s_1\rangle$};
\draw[color=black] (36.000000,0.000000) node[right] {$|s_0\rangle$};
\end{tikzpicture}
  \caption{One-bit addition circuit.
  The result of $a+b$ is stored in the two-bit number $s = s_0s_1$.} \label{circ:one-bit-adder}
\end{figure}
At the logical level, adding two bits (potentially in superposition), may be carried out using the circuit in \Cref{circ:one-bit-adder}.
Note that a classical one-bit addition requires only two bits of memory, as two bits are sufficient to store both the input bits and the output, which may be a two-bit number.
However, the resulting circuit is not reversible, as the input bits cannot be recovered from their sum.
To implement the one-bit adder on a quantum computer, a reversible classical circuit is needed.
The one-bit adder can be made reversible at the cost of requiring three bits or qubits.

\begin{figure}
  \begin{tikzpicture}[scale=1.000000,x=1pt,y=1pt]
\filldraw[color=white] (0.000000, -7.500000) rectangle (66.000000, 37.500000);
\draw[color=black] (0.000000,30.000000) -- (54.000000,30.000000);
\draw[color=black] (54.000000,29.500000) -- (66.000000,29.500000);
\draw[color=black] (54.000000,30.500000) -- (66.000000,30.500000);
\draw[color=black] (0.000000,30.000000) node[left] {$|+\rangle$};
\draw[color=black] (0.000000,15.000000) -- (54.000000,15.000000);
\draw[color=black] (54.000000,14.500000) -- (66.000000,14.500000);
\draw[color=black] (54.000000,15.500000) -- (66.000000,15.500000);
\draw[color=black] (0.000000,15.000000) node[left] {$|+\rangle$};
\draw[color=black] (0.000000,0.000000) -- (54.000000,0.000000);
\draw[color=black] (54.000000,-0.500000) -- (66.000000,-0.500000);
\draw[color=black] (54.000000,0.500000) -- (66.000000,0.500000);
\draw[color=black] (0.000000,0.000000) node[left] {$|+\rangle$};
\draw (9.000000,30.000000) -- (9.000000,0.000000);
\filldraw (9.000000, 30.000000) circle(1.500000pt);
\filldraw (9.000000, 15.000000) circle(1.500000pt);
\filldraw (9.000000, 0.000000) circle(1.500000pt);
\draw (30.000000,30.000000) -- (30.000000,15.000000);
\filldraw (30.000000, 30.000000) circle(1.500000pt);
\begin{scope}
\draw[fill=white] (30.000000, 15.000000) circle(3.000000pt);
\clip (30.000000, 15.000000) circle(3.000000pt);
\draw (27.000000, 15.000000) -- (33.000000, 15.000000);
\draw (30.000000, 12.000000) -- (30.000000, 18.000000);
\end{scope}
\begin{scope}
\draw[fill=white] (30.000000, -0.000000) +(-45.000000:8.485281pt and 8.485281pt) -- +(45.000000:8.485281pt and 8.485281pt) -- +(135.000000:8.485281pt and 8.485281pt) -- +(225.000000:8.485281pt and 8.485281pt) -- cycle;
\clip (30.000000, -0.000000) +(-45.000000:8.485281pt and 8.485281pt) -- +(45.000000:8.485281pt and 8.485281pt) -- +(135.000000:8.485281pt and 8.485281pt) -- +(225.000000:8.485281pt and 8.485281pt) -- cycle;
\draw (30.000000, -0.000000) node {$H$};
\end{scope}
\draw[fill=white] (48.000000, 24.000000) rectangle (60.000000, 36.000000);
\draw[very thin] (54.000000, 30.600000) arc (90:150:6.000000pt);
\draw[very thin] (54.000000, 30.600000) arc (90:30:6.000000pt);
\draw[->,>=stealth] (54.000000, 24.600000) -- +(80:10.392305pt);
\draw[fill=white] (48.000000, 9.000000) rectangle (60.000000, 21.000000);
\draw[very thin] (54.000000, 15.600000) arc (90:150:6.000000pt);
\draw[very thin] (54.000000, 15.600000) arc (90:30:6.000000pt);
\draw[->,>=stealth] (54.000000, 9.600000) -- +(80:10.392305pt);
\draw[fill=white] (48.000000, -6.000000) rectangle (60.000000, 6.000000);
\draw[very thin] (54.000000, 0.600000) arc (90:150:6.000000pt);
\draw[very thin] (54.000000, 0.600000) arc (90:30:6.000000pt);
\draw[->,>=stealth] (54.000000, -5.400000) -- +(80:10.392305pt);
\draw[color=black] (66.000000,30.000000) node[right] {$a$};
\draw[color=black] (66.000000,15.000000) node[right] {$s_1$};
\draw[color=black] (66.000000,0.000000) node[right] {$s_0$};
\end{tikzpicture} \caption{One-bit addition with the uniform superposition as input, preparing the `superposition of valid sums' and measuring it destructively.} \label{circ:one-bit-adder-superposition}

\end{figure}
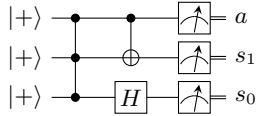

Implementing the circuit in \Cref{circ:one-bit-adder} with computational basis states as inputs would not be a good demonstration of quantum computing, as the state would not exhibit superposition or entanglement throughout the computation.
In order to remedy this, we input the state $\ket{+}\ket{+}$, thus preparing the equal superposition of all four two-bit sums, and obtaining the result of one of them at measurement time.
We also replace the Toffoli gate with a $\ccz$ gate conjugated by Hadamards, which, after simplification, gives the circuit of \Cref{circ:one-bit-adder-superposition}, which is the logical circuit we use in the remainder of this work.

In the following section, we review the error-detecting code selected for this work (the $\SICC$ colour code). We express the logical circuit in terms of simple, low-overhead fault-tolerant operations in \cref{sec:construction}.
We then implement the resulting circuit in an ion-trap quantum computer, comparing the arithmetic error rates achieved using fault-tolerant and non-fault-tolerant circuits, as well as simulations, in \cref{sec:results}.
We discuss the resource overhead required for a variety of implementations of one-bit addition in \cref{sec:resource-comparisons}, and conclude in \cref{sec:conclusion}.

\section{$\SICC$ colour code} \label{sec:colour-code}
To encode the three logical qubits necessary for one-bit addition and gain access to a transversal $\ccz$, we select the $\SICC$ colour code \cite{SmallestInterestingColour2016}.
\begin{figure}
	\resizebox{0.4\textwidth}{!}{\input{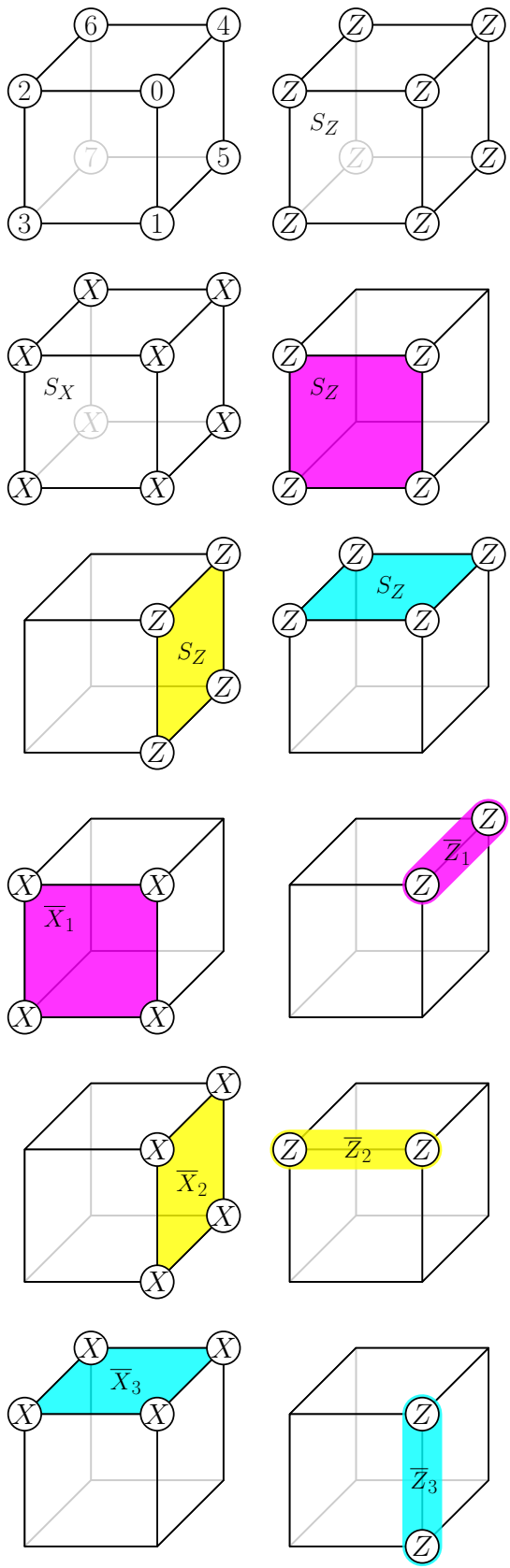}}
	\caption{Stabilizer generators and logical operators for the $\SICC$ colour code. }
	\label{fig:basic-cube}
\end{figure}
To prepare a uniform superposition of valid one-bit sums, we first fault-tolerantly prepare the $\ket{\overline{+}}^{\otimes 3}$ state.
We then perform the transversal \ccz, and a \cnot{} between qubits $1$ and $2$, then measure qubit $3$ in the $\overline{X}$ basis, and finally measure qubits $1$ and $2$ in the $\overline{Z}$ basis.
While the transversal \ccz{} of the $\SICC$ code is already well understood, the state preparation, Clifford gates and measurements used here have not been described in the prior literature to our knowledge. 
We explain their derivation below, in order of execution in the experiment.

\section{\label{sec:construction} Construction of the circuit}
\subsection{$\ket{\overline{+}}^{\otimes 3}$ Preparation}
There is a well-known procedure for fault-tolerant preparation of $\ket{\overline{+}}^{\otimes k}$ states in CSS codes of distance $d$ involving measuring $Z$ stabilisers with the $\ket{+}^{\otimes n}$ state as input. 
The first round of measurement will result in random outcomes, so multiple rounds (two for codes with $d=2$) would be necessary to detect errors. 
For the $\SICC$ code, this requires $\sim 32$ \cnot{s} and 8 measurements. 

To reduce the size of this circuit, we use the Goto circuit design technique \cite{gotoMinimizingResourceOverheads2016}, first writing out a non-fault-tolerant circuit with the minimum number of two-qubit gates, then measuring a limited set of stabilizers that detect high-weight errors resulting from error propagation through the initial circuit. 
We find the non-fault-tolerant stage of the circuit by inspection, beginning from the desired final state and using \cnot{} gates to break the state's entanglement, until arriving at a state consisting of four Bell pairs, which can be prepared fault-tolerantly using \cnot{s} on bare qubits.
We confirm that this circuit contains the minimum number of \cnot{s} using breadth-first search over an implicit graph whose vertices are canonical stabilizer states (see \cite{scheinermanImplicitGraphs2022}, \cite{QuantumCliffordJl2023}).

Gottesman-Knill simulation reveals that all high-weight propagated errors can be detected by fault-tolerant measurement of two weight-four stabilisers, $X_{1,3,4,6}$ and $Z_{1,3,4,6}$. 
This can be accomplished using an appropriately interleaved circuit designed by Reichardt \cite{reichardtFaulttolerantQuantumError2018}.
The final preparation circuit requires 18 \cnot{s} and two measurements, halving the number of relatively error-prone gates with respect to the generic technique.
\subsection{Destructive $\overline{X}$ Measurement}
Similarly to state preparation, there is also a generic protocol for measuring the logical observables of CSS codes.
In this protocol, all data qubits are measured in the $X$ or $Z$ basis, and the eigenvalues of stabilisers and logical operators in that basis are then reconstructed by calculating classical parities, allowing a final round of classical error correction to be performed.
In this way, any set composed only of tensor products of $\overline{Z}$ or
$\overline{X}$ operators can be measured, but we cannot simultaneously measure $\overline{X}_j$ and $\overline{Z}_{k \neq j}$ using this protocol.

\begin{figure}
	\resizebox{0.4\textwidth}{!}{\input{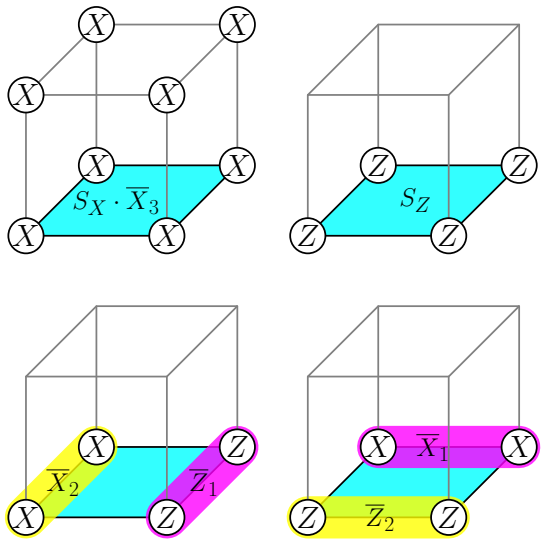}}
	\caption{Action of measuring every qubit of the top face of the cube in the $X$ basis.}
	\label{fig:destructive-X-measurement}
\end{figure}

Typically, whenever we wish to measure a subset of logical qubits in a different basis, we use an ancilla-based circuit to measure the relevant operators in a non-destructive, fault-tolerant way (similarly to stabilizer measurements) or synthesize the logical measurement by transferring the relevant subset of logical qubits to a new code block, which may subsequently be measured destructively. 
The $\SICC$ colour code allows an alternative to such a protocol; a logical $X$ operator supported on a face may be measured destructively by measuring the four qubits on that face in the $X$ basis. 
The remaining $\overline{X}$ operators are reduced to weight two, and the weights of the $\overline{Z}$ operators not supported on the measured face are preserved, leaving the remaining two logical qubits encoded in the $\nkd{4}{2}{2}$ code, see \Cref{fig:destructive-X-measurement}. 

This destructive measurement is not fault-tolerant because no stabiliser eigenvalue can be reconstructed from the measurement outputs.
To reconstruct $S_X = X^{\otimes 8}$, we use a flag-based circuit to non-destructively measure $X^{\otimes 4}$ on the opposite face and take the parity of the five measurement outputs to reconstruct the stabilizer eigenvalue.

\subsection{Logical CNOT by Permutation}
\begin{figure}
	\resizebox{0.4\textwidth}{!}{\input{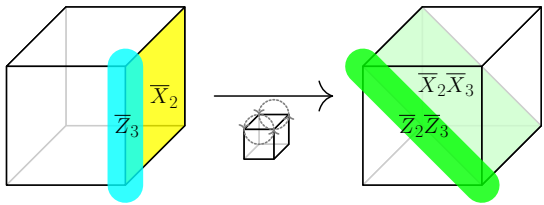}}
	\caption{Action of reflecting the top face of the cubic layout for the $\SICC$ code.
	The operators $\overline{X}_2$ and $\overline{Z}_{3}$ are mapped to $\overline{X}_2 \overline{X}_3$ and $\overline{Z}_2 \overline{Z}_3$, respectively, leaving other logical operators unaffected, effectively performing a logical \cnot{} gate.}
	\label{fig:cnot-by-permutation}
\end{figure}

With CSS codes that encode a single logical qubit, such as surface code patches, entangling operations are implemented across different code blocks, usually using transversal gates or lattice surgery \cite{terhalQuantumErrorCorrection2015}.
For codes such as the $\SICC$ code, which possess $k>1$ logical qubits, there is no general protocol for performing logical entangling operations within a single code block (though architectures which use codes with $k>1$ have been explored \cite{brun2015teleportationbased}).

However, the $\SICC$ colour code possesses a logical $\cnot$ gate which can be implemented by permuting or relabelling the physical qubits.
This is due to a spatial symmetry of the stabilizer group which the logical operators do not obey.
The QCCD architecture \cite{kielpinskiArchitectureLargescaleIontrap2002, pinoDemonstrationTrappedionQuantum2021} allows us to transport physical qubits from one area of a device to another if necessary.
As a result, the $\cnot$ gate given in \Cref{circ:one-bit-adder-superposition} can be implemented with near-unit fidelity using only transport operations, see \Cref{fig:cnot-by-permutation}.
\section{\label{sec:results} Experimental Results}
\subsection{Experimental details}
\begin{figure*}
	\resizebox{\textwidth}{!}{\input{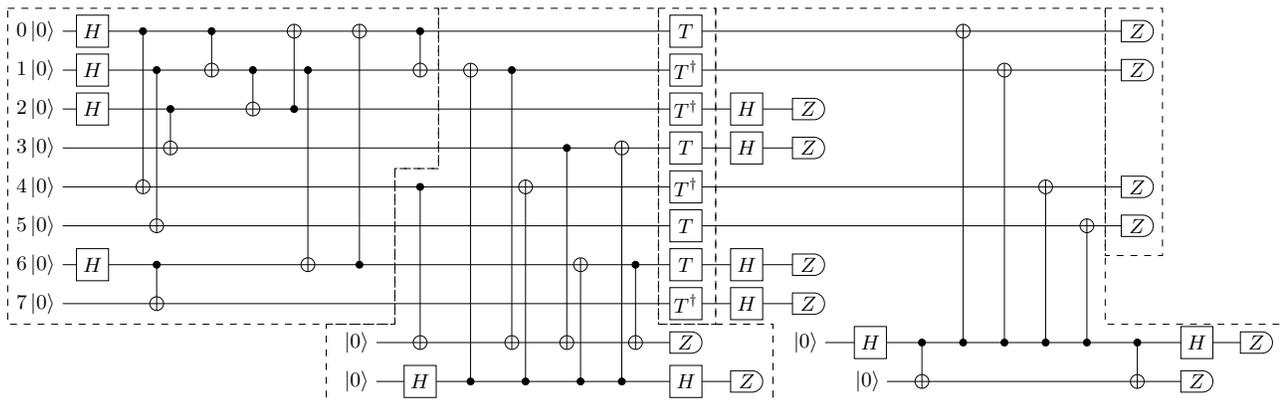}}
	\caption{Fault-tolerant implementation of one-bit addition given in \Cref{circ:one-bit-adder-superposition} as submitted to the compiler. 
	Dashed regions, from left to right: Non-fault-tolerant $\ket{\overline{+}\overline{+}\overline{+}}$ preparation, flag fault-tolerant measurement of $\overline{X}_1\overline{X}_3$ and overlapping $S_Z$, transversal \ccz{,} destructive measurement of $\overline{X}_2$, destructive measurement of $\overline{Z}_1$ and  $\overline{Z}_3$.}
	\label{fig:full_FT_circuit}
\end{figure*}

\begin{figure}
	\resizebox{0.48\textwidth}{!}{\begin{tikzpicture}[scale=1.000000,x=1pt,y=1pt]
\filldraw[color=white] (0.000000, -7.500000) rectangle (240.000000, 37.500000);
\draw[color=black] (0.000000,30.000000) -- (228.000000,30.000000);
\draw[color=black] (0.000000,30.000000) node[left] {$\ket{0}$};
\draw[color=black] (0.000000,15.000000) -- (228.000000,15.000000);
\draw[color=black] (0.000000,15.000000) node[left] {$\ket{0}$};
\draw[color=black] (0.000000,0.000000) -- (228.000000,0.000000);
\draw[color=black] (0.000000,0.000000) node[left] {$\ket{0}$};
\begin{scope}
\draw[fill=white] (12.000000, 30.000000) +(-45.000000:8.485281pt and 8.485281pt) -- +(45.000000:8.485281pt and 8.485281pt) -- +(135.000000:8.485281pt and 8.485281pt) -- +(225.000000:8.485281pt and 8.485281pt) -- cycle;
\clip (12.000000, 30.000000) +(-45.000000:8.485281pt and 8.485281pt) -- +(45.000000:8.485281pt and 8.485281pt) -- +(135.000000:8.485281pt and 8.485281pt) -- +(225.000000:8.485281pt and 8.485281pt) -- cycle;
\draw (12.000000, 30.000000) node {$H$};
\end{scope}
\begin{scope}
\draw[fill=white] (12.000000, 15.000000) +(-45.000000:8.485281pt and 8.485281pt) -- +(45.000000:8.485281pt and 8.485281pt) -- +(135.000000:8.485281pt and 8.485281pt) -- +(225.000000:8.485281pt and 8.485281pt) -- cycle;
\clip (12.000000, 15.000000) +(-45.000000:8.485281pt and 8.485281pt) -- +(45.000000:8.485281pt and 8.485281pt) -- +(135.000000:8.485281pt and 8.485281pt) -- +(225.000000:8.485281pt and 8.485281pt) -- cycle;
\draw (12.000000, 15.000000) node {$H$};
\end{scope}
\begin{scope}
\draw[fill=white] (12.000000, -0.000000) +(-45.000000:8.485281pt and 8.485281pt) -- +(45.000000:8.485281pt and 8.485281pt) -- +(135.000000:8.485281pt and 8.485281pt) -- +(225.000000:8.485281pt and 8.485281pt) -- cycle;
\clip (12.000000, -0.000000) +(-45.000000:8.485281pt and 8.485281pt) -- +(45.000000:8.485281pt and 8.485281pt) -- +(135.000000:8.485281pt and 8.485281pt) -- +(225.000000:8.485281pt and 8.485281pt) -- cycle;
\draw (12.000000, -0.000000) node {$H$};
\end{scope}
\draw (33.000000,15.000000) -- (33.000000,0.000000);
\begin{scope}
\draw[fill=white] (33.000000, 0.000000) circle(3.000000pt);
\clip (33.000000, 0.000000) circle(3.000000pt);
\draw (30.000000, 0.000000) -- (36.000000, 0.000000);
\draw (33.000000, -3.000000) -- (33.000000, 3.000000);
\end{scope}
\filldraw (33.000000, 15.000000) circle(1.500000pt);
\begin{scope}
\draw[fill=white] (54.000000, -0.000000) +(-45.000000:8.485281pt and 8.485281pt) -- +(45.000000:8.485281pt and 8.485281pt) -- +(135.000000:8.485281pt and 8.485281pt) -- +(225.000000:8.485281pt and 8.485281pt) -- cycle;
\clip (54.000000, -0.000000) +(-45.000000:8.485281pt and 8.485281pt) -- +(45.000000:8.485281pt and 8.485281pt) -- +(135.000000:8.485281pt and 8.485281pt) -- +(225.000000:8.485281pt and 8.485281pt) -- cycle;
\draw (54.000000, -0.000000) node {$T^{\dagger}$};
\end{scope}
\draw (75.000000,30.000000) -- (75.000000,0.000000);
\begin{scope}
\draw[fill=white] (75.000000, 0.000000) circle(3.000000pt);
\clip (75.000000, 0.000000) circle(3.000000pt);
\draw (72.000000, 0.000000) -- (78.000000, 0.000000);
\draw (75.000000, -3.000000) -- (75.000000, 3.000000);
\end{scope}
\filldraw (75.000000, 30.000000) circle(1.500000pt);
\begin{scope}
\draw[fill=white] (96.000000, -0.000000) +(-45.000000:8.485281pt and 8.485281pt) -- +(45.000000:8.485281pt and 8.485281pt) -- +(135.000000:8.485281pt and 8.485281pt) -- +(225.000000:8.485281pt and 8.485281pt) -- cycle;
\clip (96.000000, -0.000000) +(-45.000000:8.485281pt and 8.485281pt) -- +(45.000000:8.485281pt and 8.485281pt) -- +(135.000000:8.485281pt and 8.485281pt) -- +(225.000000:8.485281pt and 8.485281pt) -- cycle;
\draw (96.000000, -0.000000) node {$T$};
\end{scope}
\draw (117.000000,15.000000) -- (117.000000,0.000000);
\begin{scope}
\draw[fill=white] (117.000000, 0.000000) circle(3.000000pt);
\clip (117.000000, 0.000000) circle(3.000000pt);
\draw (114.000000, 0.000000) -- (120.000000, 0.000000);
\draw (117.000000, -3.000000) -- (117.000000, 3.000000);
\end{scope}
\filldraw (117.000000, 15.000000) circle(1.500000pt);
\begin{scope}
\draw[fill=white] (138.000000, -0.000000) +(-45.000000:8.485281pt and 8.485281pt) -- +(45.000000:8.485281pt and 8.485281pt) -- +(135.000000:8.485281pt and 8.485281pt) -- +(225.000000:8.485281pt and 8.485281pt) -- cycle;
\clip (138.000000, -0.000000) +(-45.000000:8.485281pt and 8.485281pt) -- +(45.000000:8.485281pt and 8.485281pt) -- +(135.000000:8.485281pt and 8.485281pt) -- +(225.000000:8.485281pt and 8.485281pt) -- cycle;
\draw (138.000000, -0.000000) node {$T^{\dagger}$};
\end{scope}
\begin{scope}
\draw[fill=white] (138.000000, 15.000000) +(-45.000000:8.485281pt and 8.485281pt) -- +(45.000000:8.485281pt and 8.485281pt) -- +(135.000000:8.485281pt and 8.485281pt) -- +(225.000000:8.485281pt and 8.485281pt) -- cycle;
\clip (138.000000, 15.000000) +(-45.000000:8.485281pt and 8.485281pt) -- +(45.000000:8.485281pt and 8.485281pt) -- +(135.000000:8.485281pt and 8.485281pt) -- +(225.000000:8.485281pt and 8.485281pt) -- cycle;
\draw (138.000000, 15.000000) node {$T$};
\end{scope}
\draw (159.000000,30.000000) -- (159.000000,0.000000);
\begin{scope}
\draw[fill=white] (159.000000, 0.000000) circle(3.000000pt);
\clip (159.000000, 0.000000) circle(3.000000pt);
\draw (156.000000, 0.000000) -- (162.000000, 0.000000);
\draw (159.000000, -3.000000) -- (159.000000, 3.000000);
\end{scope}
\filldraw (159.000000, 30.000000) circle(1.500000pt);
\begin{scope}
\draw[fill=white] (180.000000, -0.000000) +(-45.000000:8.485281pt and 8.485281pt) -- +(45.000000:8.485281pt and 8.485281pt) -- +(135.000000:8.485281pt and 8.485281pt) -- +(225.000000:8.485281pt and 8.485281pt) -- cycle;
\clip (180.000000, -0.000000) +(-45.000000:8.485281pt and 8.485281pt) -- +(45.000000:8.485281pt and 8.485281pt) -- +(135.000000:8.485281pt and 8.485281pt) -- +(225.000000:8.485281pt and 8.485281pt) -- cycle;
\draw (180.000000, -0.000000) node {$T$};
\end{scope}
\draw (180.000000,30.000000) -- (180.000000,15.000000);
\begin{scope}
\draw[fill=white] (180.000000, 15.000000) circle(3.000000pt);
\clip (180.000000, 15.000000) circle(3.000000pt);
\draw (177.000000, 15.000000) -- (183.000000, 15.000000);
\draw (180.000000, 12.000000) -- (180.000000, 18.000000);
\end{scope}
\filldraw (180.000000, 30.000000) circle(1.500000pt);
\begin{scope}
\draw[fill=white] (204.000000, 30.000000) +(-45.000000:8.485281pt and 8.485281pt) -- +(45.000000:8.485281pt and 8.485281pt) -- +(135.000000:8.485281pt and 8.485281pt) -- +(225.000000:8.485281pt and 8.485281pt) -- cycle;
\clip (204.000000, 30.000000) +(-45.000000:8.485281pt and 8.485281pt) -- +(45.000000:8.485281pt and 8.485281pt) -- +(135.000000:8.485281pt and 8.485281pt) -- +(225.000000:8.485281pt and 8.485281pt) -- cycle;
\draw (204.000000, 30.000000) node {$T$};
\end{scope}
\begin{scope}
\draw[fill=white] (204.000000, 15.000000) +(-45.000000:8.485281pt and 8.485281pt) -- +(45.000000:8.485281pt and 8.485281pt) -- +(135.000000:8.485281pt and 8.485281pt) -- +(225.000000:8.485281pt and 8.485281pt) -- cycle;
\clip (204.000000, 15.000000) +(-45.000000:8.485281pt and 8.485281pt) -- +(45.000000:8.485281pt and 8.485281pt) -- +(135.000000:8.485281pt and 8.485281pt) -- +(225.000000:8.485281pt and 8.485281pt) -- cycle;
\draw (204.000000, 15.000000) node {$T^{\dagger}$};
\end{scope}
\begin{scope}
\draw[fill=white] (204.000000, -0.000000) +(-45.000000:8.485281pt and 8.485281pt) -- +(45.000000:8.485281pt and 8.485281pt) -- +(135.000000:8.485281pt and 8.485281pt) -- +(225.000000:8.485281pt and 8.485281pt) -- cycle;
\clip (204.000000, -0.000000) +(-45.000000:8.485281pt and 8.485281pt) -- +(45.000000:8.485281pt and 8.485281pt) -- +(135.000000:8.485281pt and 8.485281pt) -- +(225.000000:8.485281pt and 8.485281pt) -- cycle;
\draw (204.000000, -0.000000) node {$H$};
\end{scope}
\draw[fill=white] (218.000000, 30.000000) -- (222.000000,34.000000) -- (234.000000,34.000000) -- (234.000000, 26.000000) -- (222.000000, 26.000000) -- cycle;
\draw (228.000000, 30.000000) node {$Z$};
\draw[fill=white] (218.000000, 15.000000) -- (222.000000,19.000000) -- (234.000000,19.000000) -- (234.000000, 11.000000) -- (222.000000, 11.000000) -- cycle;
\draw (228.000000, 15.000000) node {$Z$};
\draw[fill=white] (218.000000, 0.000000) -- (222.000000,4.000000) -- (234.000000,4.000000) -- (234.000000, -4.000000) -- (222.000000, -4.000000) -- cycle;
\draw (228.000000, 0.000000) node {$Z$};
\end{tikzpicture}}
	\caption{Non-fault-tolerant circuit as submitted to compiler.}
	\label{fig:non_FT_circuit}
\end{figure}
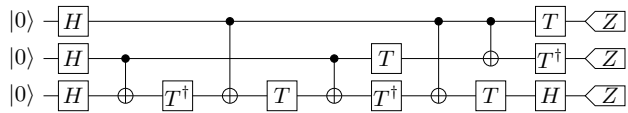

The circuit given in \Cref{fig:full_FT_circuit}, which is a fault-tolerant implementation of the one-bit adder circuit, was submitted to both the Quantinuum H1-1 quantum computer and the Quantinuum H1-1E emulator.
Upon submission, a compiler transforms the circuit into one corresponding to the native gate set of the quantum computer.
For comparison, a non-fault tolerant circuit acting on three physical qubits, given in \Cref{fig:non_FT_circuit}, was also submitted to the device.
The resulting circuits were executed 10,000 times on the quantum computer and 100,000 times on the emulator, and the results presented in \Cref{tab:error_rates}.
\begin{table*}[tb]
\begin{tabular}{| c | c | c | c | c | }
  \hline
  \multirow{2}{3cm}{Shot frequencies} & \multicolumn{2}{|c|}{Non-fault-tolerant} & \multicolumn{2}{|c|}{Fault-tolerant} \\
  \cline{2-5}
   & Device & Emulator & Device & Emulator \\
  \hline
  \cellcolor{green!30!white} 000 & 25.18\% & 24.64\% & 24.49\% & 24.98\% \\
  \cellcolor{red!30!white} 001 & 0.14\% & 0.30\% & 0.03\% & 0.01\%\\
  \cellcolor{green!30!white} 010 & 25.23\% & 24.86\% & 25.52\% & 25.15\% \\
  \cellcolor{red!30!white} 011 & 0.07\% & 0.19\% & 0.02\% & 0.008\% \\
  \cellcolor{red!30!white} 100 & 0.41\% & 0.46\% & 0.02\% & 0.012\% \\
  \cellcolor{green!30!white} 101 & 24.52\% & 24.58\% & 24.75\% & 24.87\% \\
  \cellcolor{green!30!white} 110 & 24.14\% & 24.70\% & 25.15\% & 24.95\% \\
  \cellcolor{red!30!white} 111 & 0.31\% & 0.27\% & 0.01\% & 0.01\% \\
  \hline
  Shot total & 10000 & 100000 & 8998 & 88537 \\
  \hline
  Arithmetic error rate & 0.95±0.19\% & 1.22±0.07\% & 0.11±0.07\% & 0.04±0.014\% \\
  \hline
\end{tabular}
  \caption{Results of evaluating fault-tolerant and non-fault-tolerant circuits on both the Quantinuum H1-1 quantum computer and emulator.
  Only those shots that pass post-selection are used to estimate the arithmetic error rate of the fault-tolerant circuit.
  Rows in red correspond to arithmetic errors, while the rows in green correspond to valid one-bit sums.}
\label{tab:error_rates}
\end{table*}

The implementations of logical operations discussed in the previous section allow us to write a fault-tolerant implementation of one-bit addition in the $\SICC$ colour code using 24 \cnot{s} and 12 measurements \footnote{Single-qubit operations are not counted for the purpose of estimating the final logical error rate, due to their much lower error rates.}, see \Cref{fig:full_FT_circuit}. 
For comparison, decomposing the one-qubit adder into \cnot{s} and single-qubit gates on bare qubits results in five \cnot{s} and three measurements, see \Cref{fig:non_FT_circuit}. 

To completely characterize the logical errors which occur in a fault-tolerant one-qubit addition, performing three-qubit logical process tomography would be necessary.
However, process tomography results in prohibitively high sampling overhead and introduces the challenge of distinguishing state preparation and measurement (SPAM) errors from those occurring in the unitary of interest.
For these reasons, we instead execute a complete protocol consisting of state preparation, unitary operations and measurements, and calculate an operationally-defined error rate which is affected by all steps of the process.
At the end of each summation, we measure the register in the computational basis, at which point we learn classical values for the one-bit number $a$, and the two-bit number $s$. If
\begin{equation}
(s = 3) \lor ((s = 2) \land (a = 0)) \lor ((s = 0) \land (a = 1)),
\end{equation}
we can infer that a logical error has occurred. We call these events \emph{arithmetic errors} and compare the rates at which they occur in \cref{tab:error_rates}.

In order to evaluate the effectiveness of the arithmetic error rate as a measure of the true error rate, we attempt to estimate the true error rate.
This cannot be done directly from the data presented in \Cref{tab:error_rates}.
Therefore, we carry out a density-matrix-based simulation of the circuit, using a noise model motivated by that of the Quantinuum H1-1 emulator, and attempt to quantify the effect of noise on the state obtained at the end of the circuit.
We obtain an arithmetic error rate of 0.39\%, which is of comparable magnitude to that observed in \Cref{tab:error_rates}.
The fidelity of the simulated state with the ideal error-free state 99.7\%, which implies that the use of the arithmetic error rate does not conceal significant logical errors in other bases.
\section{Comparison With Planar Architectures} \label{sec:resource-comparisons}
To understand the overhead reduction achieved in this implementation, we investigate a surface-code-based implementation of a one-bit addition for surface code patches with distance $d=2$.
Similar to \cite{litinskiGameSurfaceCodes2019}, logical qubits are realised in independent code blocks, logical Clifford gates are implemented using lattice surgery, and the logical Hadamard is propagated forward, resulting in an $\overline{X}$ measurement on the third qubit at the end of the circuit.
We can lower the overhead of this implementation by generating the state $\ccz \ket{\overline{+}\overline{+}\overline{+}}$ using a magic state factory \cite{Gidney2019efficientmagicstate}, and computing with it directly, rather than using it for gate teleportation.
At distance $d=2$, the factory can detect any single-qubit error during the production of the $\ket{\ccz}$ state.

The overhead of this implementation is dominated by the $\ccz$ magic state factory.
It requires 18 surface code patches to implement, which at distance $d=2$ implies $\sim 18 \cdot 2d^2 = 144$ physical qubits.
This figure is over an order of magnitude higher than that of the $\SICC$ code implementation discussed above and is too large to be executed on a Quantinuum H-series device at time of writing.

\begin{figure*}
\hfill
\includegraphics[width=0.2\textwidth]{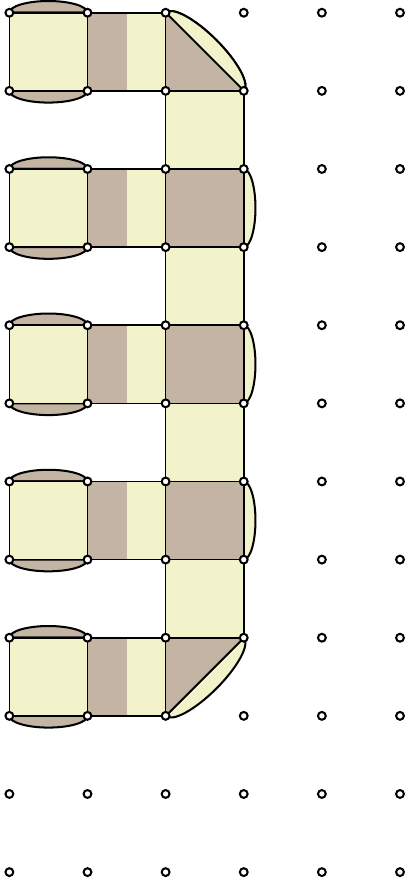}
\hfill
\includegraphics[width=0.2\textwidth]{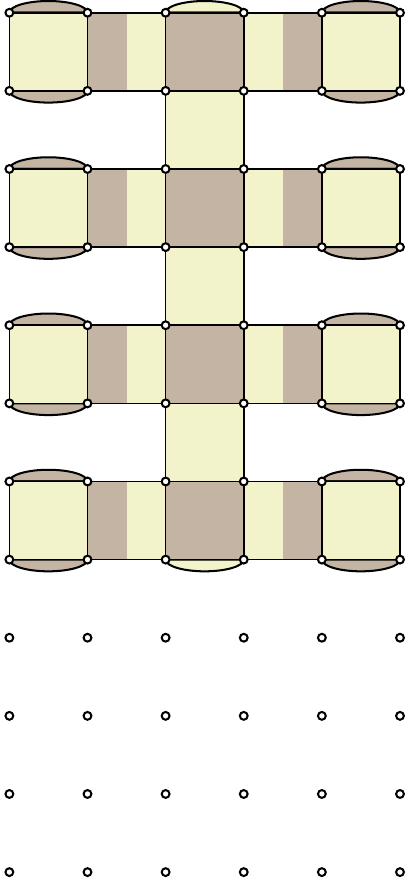}
\hfill
\includegraphics[width=0.2\textwidth]{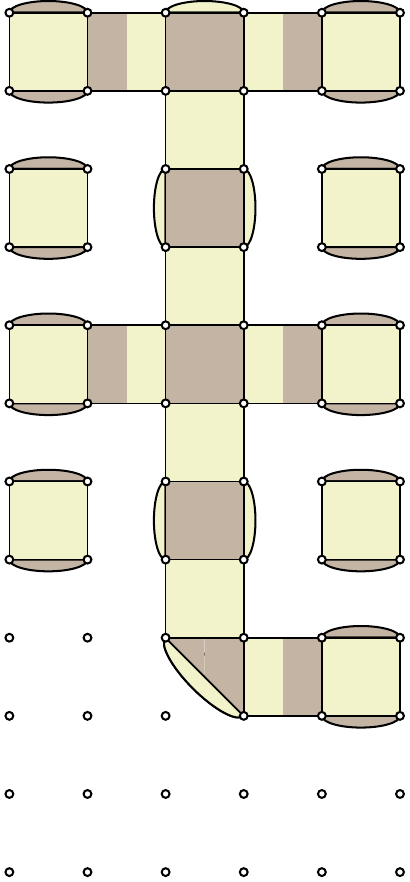}
\hfill
\includegraphics[width=0.2\textwidth]{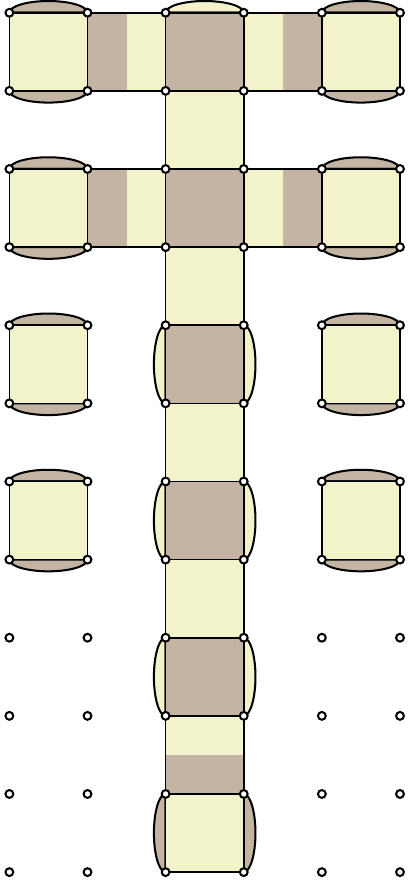}
\hfill \\
\vspace{12pt}
\includegraphics[width=0.6\textwidth]{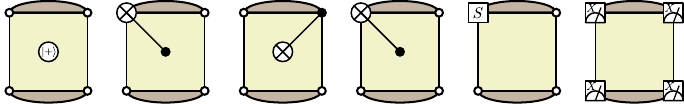}
\caption{Schedule for simulating a \ccz{} state factory \cite{Gidney2019efficientmagicstate} using lattice surgery with distance-2 surface codes.
Top: multi-qubit logical $\overline{X}$ measurements performed in series. 
Bottom: Destructive distance-1 measurement in the $\overline{S} \overline{X}$ basis, applied to the eight surface codes used as logical ancillas. 
Note that, in a genuine magic state factory, $T$ and $T\ct$ gates would take the place of the $S$ gates used here, which we use here for ease of simulation.  
Not shown: Pauli corrections applied to the three output logical qubits depending on results of destructive measurement (see \cite{Gidney2019efficientmagicstate}).}
\end{figure*}

The relative logical error rates resulting from the use of such different state factories can be estimated by counting pairs of faults that result in logical errors (larger sets of independent faults being much less likely). 
We carry this estimation out using \texttt{QuantumClifford.jl}
\cite{QuantumCliffordJl2023}, resulting in 84873 fault pairs for the $d=2$ surface code, and 1116 for the $\SICC$ code.
While a more detailed comparison would not apply to devices of either architecture, we can see that the number of malicious pairs is far greater than the number of malicious pairs for the $\SICC$-code-based implementation, which suggests that higher code distance (and greater overhead) would be necessary to match the logical error rate obtained in this work. 

\begin{figure}
\centering
\resizebox{0.35\textwidth}{!}{\input{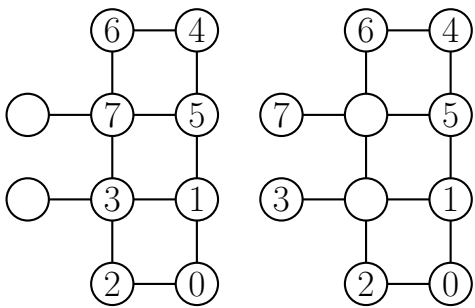}}
\caption{Small planar qubit layouts that facilitate fault-tolerant one-bit addition.
Left: layout for non-fault-tolerant state preparation. 
Right: layout for subsequent flag-based measurement of $X_{1,2,5,6}$/$Z_{1,2,5,6}$ and $X_{1,3,5,7}$. }
\label{fig:planar_832_layouts}
\end{figure}

By contrast, implementing one-qubit addition using the $\SICC$ colour code on a device with square-lattice connectivity can be accomplished with moderate overhead using the qubit layouts in \Cref{fig:planar_832_layouts}. 
The initial layout is used for non-fault-tolerant state preparation, and four \cnot{s} are then required to change the layout so that the remainder of the experiment can be carried out. 
After non-fault-tolerant state preparation, the stabilizer measurement requires an additional six \cnot{s}, since the \swap{} gates must be decomposed into \cnot{s} rather than replaced with transport operations.
While these additional \cnot{s} represent a significant increase in the overall size of the circuit without increasing its ability to tolerate errors, the induced overhead is not as significant as implementing the computation with surface-code-based logical qubits. 

\begin{figure}
\centering
\resizebox{0.35\textwidth}{!}{\begin{tikzpicture}[scale=1.000000,x=1pt,y=1pt]
\filldraw[color=white] (0.000000, -7.500000) rectangle (108.000000, 112.500000);
\draw[color=black] (0.000000,105.000000) -- (108.000000,105.000000);
\draw[color=black] (0.000000,105.000000) node[left] {${0 \ket{0}}$};
\draw[color=black] (0.000000,90.000000) -- (108.000000,90.000000);
\draw[color=black] (0.000000,90.000000) node[left] {${1 \ket{0}}$};
\draw[color=black] (0.000000,75.000000) -- (108.000000,75.000000);
\draw[color=black] (0.000000,75.000000) node[left] {${2 \ket{0}}$};
\draw[color=black] (0.000000,60.000000) -- (108.000000,60.000000);
\draw[color=black] (0.000000,60.000000) node[left] {${3 \ket{0}}$};
\draw[color=black] (0.000000,45.000000) -- (108.000000,45.000000);
\draw[color=black] (0.000000,45.000000) node[left] {${4 \ket{0}}$};
\draw[color=black] (0.000000,30.000000) -- (108.000000,30.000000);
\draw[color=black] (0.000000,30.000000) node[left] {${5 \ket{0}}$};
\draw[color=black] (0.000000,15.000000) -- (108.000000,15.000000);
\draw[color=black] (0.000000,15.000000) node[left] {${6 \ket{0}}$};
\draw[color=black] (0.000000,0.000000) -- (108.000000,0.000000);
\draw[color=black] (0.000000,0.000000) node[left] {${7 \ket{0}}$};
\begin{scope}
\draw[fill=white] (12.000000, 105.000000) +(-45.000000:8.485281pt and 8.485281pt) -- +(45.000000:8.485281pt and 8.485281pt) -- +(135.000000:8.485281pt and 8.485281pt) -- +(225.000000:8.485281pt and 8.485281pt) -- cycle;
\clip (12.000000, 105.000000) +(-45.000000:8.485281pt and 8.485281pt) -- +(45.000000:8.485281pt and 8.485281pt) -- +(135.000000:8.485281pt and 8.485281pt) -- +(225.000000:8.485281pt and 8.485281pt) -- cycle;
\draw (12.000000, 105.000000) node {$H$};
\end{scope}
\begin{scope}
\draw[fill=white] (12.000000, 60.000000) +(-45.000000:8.485281pt and 8.485281pt) -- +(45.000000:8.485281pt and 8.485281pt) -- +(135.000000:8.485281pt and 8.485281pt) -- +(225.000000:8.485281pt and 8.485281pt) -- cycle;
\clip (12.000000, 60.000000) +(-45.000000:8.485281pt and 8.485281pt) -- +(45.000000:8.485281pt and 8.485281pt) -- +(135.000000:8.485281pt and 8.485281pt) -- +(225.000000:8.485281pt and 8.485281pt) -- cycle;
\draw (12.000000, 60.000000) node {$H$};
\end{scope}
\begin{scope}
\draw[fill=white] (12.000000, -0.000000) +(-45.000000:8.485281pt and 8.485281pt) -- +(45.000000:8.485281pt and 8.485281pt) -- +(135.000000:8.485281pt and 8.485281pt) -- +(225.000000:8.485281pt and 8.485281pt) -- cycle;
\clip (12.000000, -0.000000) +(-45.000000:8.485281pt and 8.485281pt) -- +(45.000000:8.485281pt and 8.485281pt) -- +(135.000000:8.485281pt and 8.485281pt) -- +(225.000000:8.485281pt and 8.485281pt) -- cycle;
\draw (12.000000, -0.000000) node {$H$};
\end{scope}
\begin{scope}
\draw[fill=white] (12.000000, 15.000000) +(-45.000000:8.485281pt and 8.485281pt) -- +(45.000000:8.485281pt and 8.485281pt) -- +(135.000000:8.485281pt and 8.485281pt) -- +(225.000000:8.485281pt and 8.485281pt) -- cycle;
\clip (12.000000, 15.000000) +(-45.000000:8.485281pt and 8.485281pt) -- +(45.000000:8.485281pt and 8.485281pt) -- +(135.000000:8.485281pt and 8.485281pt) -- +(225.000000:8.485281pt and 8.485281pt) -- cycle;
\draw (12.000000, 15.000000) node {$H$};
\end{scope}
\draw (33.000000,105.000000) -- (33.000000,90.000000);
\begin{scope}
\draw[fill=white] (33.000000, 90.000000) circle(3.000000pt);
\clip (33.000000, 90.000000) circle(3.000000pt);
\draw (30.000000, 90.000000) -- (36.000000, 90.000000);
\draw (33.000000, 87.000000) -- (33.000000, 93.000000);
\end{scope}
\filldraw (33.000000, 105.000000) circle(1.500000pt);
\draw (33.000000,75.000000) -- (33.000000,60.000000);
\begin{scope}
\draw[fill=white] (33.000000, 75.000000) circle(3.000000pt);
\clip (33.000000, 75.000000) circle(3.000000pt);
\draw (30.000000, 75.000000) -- (36.000000, 75.000000);
\draw (33.000000, 72.000000) -- (33.000000, 78.000000);
\end{scope}
\filldraw (33.000000, 60.000000) circle(1.500000pt);
\draw (51.000000,90.000000) -- (51.000000,30.000000);
\begin{scope}
\draw[fill=white] (51.000000, 30.000000) circle(3.000000pt);
\clip (51.000000, 30.000000) circle(3.000000pt);
\draw (48.000000, 30.000000) -- (54.000000, 30.000000);
\draw (51.000000, 27.000000) -- (51.000000, 33.000000);
\end{scope}
\filldraw (51.000000, 90.000000) circle(1.500000pt);
\draw (57.000000,60.000000) -- (57.000000,0.000000);
\begin{scope}
\draw[fill=white] (57.000000, 60.000000) circle(3.000000pt);
\clip (57.000000, 60.000000) circle(3.000000pt);
\draw (54.000000, 60.000000) -- (60.000000, 60.000000);
\draw (57.000000, 57.000000) -- (57.000000, 63.000000);
\end{scope}
\filldraw (57.000000, 0.000000) circle(1.500000pt);
\draw (57.000000,105.000000) -- (57.000000,75.000000);
\begin{scope}
\draw[fill=white] (57.000000, 105.000000) circle(3.000000pt);
\clip (57.000000, 105.000000) circle(3.000000pt);
\draw (54.000000, 105.000000) -- (60.000000, 105.000000);
\draw (57.000000, 102.000000) -- (57.000000, 108.000000);
\end{scope}
\filldraw (57.000000, 75.000000) circle(1.500000pt);
\draw (75.000000,45.000000) -- (75.000000,30.000000);
\begin{scope}
\draw[fill=white] (75.000000, 45.000000) circle(3.000000pt);
\clip (75.000000, 45.000000) circle(3.000000pt);
\draw (72.000000, 45.000000) -- (78.000000, 45.000000);
\draw (75.000000, 42.000000) -- (75.000000, 48.000000);
\end{scope}
\filldraw (75.000000, 30.000000) circle(1.500000pt);
\draw (75.000000,15.000000) -- (75.000000,0.000000);
\begin{scope}
\draw[fill=white] (75.000000, 0.000000) circle(3.000000pt);
\clip (75.000000, 0.000000) circle(3.000000pt);
\draw (72.000000, 0.000000) -- (78.000000, 0.000000);
\draw (75.000000, -3.000000) -- (75.000000, 3.000000);
\end{scope}
\filldraw (75.000000, 15.000000) circle(1.500000pt);
\draw (75.000000,90.000000) -- (75.000000,60.000000);
\begin{scope}
\draw[fill=white] (75.000000, 90.000000) circle(3.000000pt);
\clip (75.000000, 90.000000) circle(3.000000pt);
\draw (72.000000, 90.000000) -- (78.000000, 90.000000);
\draw (75.000000, 87.000000) -- (75.000000, 93.000000);
\end{scope}
\filldraw (75.000000, 60.000000) circle(1.500000pt);
\draw (93.000000,30.000000) -- (93.000000,0.000000);
\begin{scope}
\draw[fill=white] (93.000000, 30.000000) circle(3.000000pt);
\clip (93.000000, 30.000000) circle(3.000000pt);
\draw (90.000000, 30.000000) -- (96.000000, 30.000000);
\draw (93.000000, 27.000000) -- (93.000000, 33.000000);
\end{scope}
\filldraw (93.000000, 0.000000) circle(1.500000pt);
\draw (99.000000,45.000000) -- (99.000000,15.000000);
\begin{scope}
\draw[fill=white] (99.000000, 45.000000) circle(3.000000pt);
\clip (99.000000, 45.000000) circle(3.000000pt);
\draw (96.000000, 45.000000) -- (102.000000, 45.000000);
\draw (99.000000, 42.000000) -- (99.000000, 48.000000);
\end{scope}
\filldraw (99.000000, 15.000000) circle(1.500000pt);
\end{tikzpicture}}
\caption{Modified non-fault-tolerant $\ket{\overline{+}\overline{+}\overline{+}}$ preparation circuit, using the minimal number of \cnot{s} and qubits connected in a square lattice.}
\end{figure}
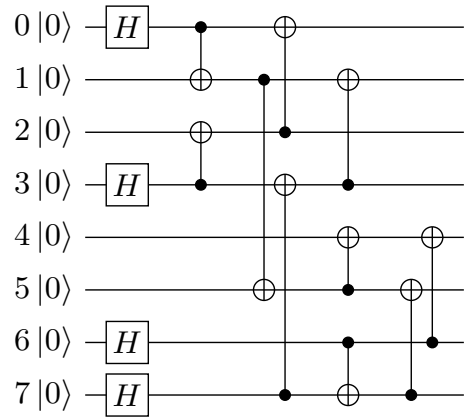
\section{Discussion \& Conclusion} \label{sec:conclusion}
The fault-tolerant implementation of one-bit addition demonstrates the combined effect of transversal non-Clifford gates, logical Cliffords by permutation, post-selected state preparation, and omission of error-correction gadgets in a fault-tolerant computation.
While we cannot expect each of these techniques to result in the same logical error rate reduction in all fault-tolerant computations, we believe that each of them will contribute to lower logical error rates and overheads in some future fault-tolerant computations. 
This result also highlights the peculiar `inversion of difficulty' in fault-tolerant quantum computing. 
That is, the operations that induce the most error at the physical level (\cnot{} and \ccz) can be carried out fault-tolerantly using high-fidelity transport and transversal single-qubit gates.
In contrast, state preparation (which is comparatively simple and reliable at the physical level) comprises most of the fault-tolerant circuit, causing a large fraction of logical errors.

This demonstration also highlights several open problems to address in future work. 
For instance, this experiment uses post-selection rather than correction and achieves a low conditional probability of error with moderate ($\sim 10\%$) post-selection overhead. 
Full post-selection (i.e. accepting the output only if no syndrome is observed) on every fault-tolerant gadget would result in an exponentially decaying acceptance probability.
Still, the effect of partial post-selection (accepting syndromes that indicate an error with weight $w \ll \nicefrac{d}{2}$ on selected gadgets within a larger algorithm) has yet to be explored.

In addition, the placement of QEC gadgets between every adjacent pair of fault-tolerant operations is sufficient to prove the existence of thresholds in the ExRec formalism \cite{aliferisQuantumAccuracyThreshold2005}.
However, it is not necessary to do this to make a circuit fault-tolerant.
Omitting QEC gadgets (or using partial QEC gadgets, as in \cref{app:422}) between consecutive fault-tolerant operations can reduce logical error rates and overheads in a wide variety of protocols.

Finally, we note that the fault-tolerant protocol we have developed leverages all-to-all connectivity of the physical qubits, and exhibits relatively high pseudothresholds for certain quantum computing tasks, making it particularly suitable for other platforms with high qubit connectivity, such as neutral atoms \cite{Xu2023, Evered2023} and NV-based networks \cite{Wang2023, Hermans2022}.
A similar idea has recently been proposed to implement quantum low-density parity-check codes (qLDPC) using reconfigurable atom arrays \cite{Xu2023}.
By using the product structure inherent in many qLDPC codes to implement non-local syndrome extraction circuits via atom rearrangement, effectively constant overhead in practically relevant regimes can be achieved \cite{Xu2023}.

\begin{acknowledgments}
The authors thank Andrew Landahl for inspiring this project, as well as Ross Duncan, Chris Self, Ciar\'an Ryan-Anderson, and David Hayes for helpful comments and insightful discussions during the preparation of this work, and Karl Mayer for illuminating discussions regarding the noise model used on the Quantinuum H1-1 emulator. 
This publication is part of the QuTech NWO funding 2020-2024 - Part I ``Fundamental Research'' with project number 601.QT.001-1, which is financed by the Dutch Research Council (NWO).
Y. Wang would like to acknowledge the funding support from BMBF (Grant No. 16KIS1590K).

\emph{Note added}: We would like to bring the reader's attention to a related work by Menendez at al., ``Implementing fault-tolerant non-Clifford gates using the [[8, 3, 2]] color code'', which appears in the same arXiv posting.
\end{acknowledgments}

\bibliography{references}

\appendix
\section{Logical Hadamard gates on the $\SICC$ code}
\label{app:422}
The selection of the $\SICC$ code is motivated by its transversal logical $\ccz$ gate, which is desirable for achieving fault-tolerant universal quantum computation.
The set $\left \lbrace H,\, \ccz \right \rbrace$ is known to be universal \cite{HandCCZareUniversal}, but cannot be performed transversally in a distance-2 code \cite{eastinKnill}.
In the one-bit addition circuit, the logical Hadamard gates can be assimilated into adjacent state preparation or measurement steps. However, we cannot expect this to be the case for an arbitrary circuit.
In general, it is possible to implement a Hadamard gate, or other Clifford gate, by fault-tolerantly preparing a Pauli eigenstate, and non-destructively projecting onto eigenspaces of logical Pauli operators, similarly to lattice surgery \cite{Horsman_2012}.

Such a construction is likely to outperform state-of-the-art magic state distillation methods \cite{UpToDateMSDOverheads} in terms of overhead for a wide variety of codes. 
However, there exists an alternative construction for the $\SICC$ code, using fewer qubits and gates overall. 
We describe this low-overhead fault-tolerant implementation of the logical Hadamard in the remainder of this appendix.

\begin{table}
    \centering
    \begin{tabular}{|cc|cccccccc|cccc|}
	\hline
	&  & $X$ & $X$ & $X$ & $X$ & $X$ & $X$ & $X$ & $X$ & $I$ & $I$ & $I$ & $I$ \\
	&  & $Z$ & $Z$ & $Z$ & $Z$ & $Z$ & $Z$ & $Z$ & $Z$ & $I$ & $I$ & $I$ & $I$ \\
	$\SICC$ & \text{Stabilizers} & $Z$ & $Z$ & $Z$ & $Z$ & $I$ & $I$ & $I$ & $I$ & $I$ & $I$ & $I$ & $I$ \\
	&  & $Z$ & $Z$ & $I$ & $I$ & $Z$ & $Z$ & $I$ & $I$ & $I$ & $I$ & $I$ & $I$ \\
	&  & $Z$ & $I$ & $Z$ & $I$ & $Z$ & $I$ & $Z$ & $I$ & $I$ & $I$ & $I$ & $I$ \\
	  \hline
	\text{$\nkd{4}{2}{2}$} & \text{Stabilizers} & $I$ & $I$ & $I$ & $I$ & $I$ & $I$ & $I$ & $I$ & $X$ & $X$ & $X$ & $X$ \\
	  &  & $I$ & $I$ & $I$ & $I$ & $I$ & $I$ & $I$ & $I$ & $Z$ & $Z$ & $Z$ & $Z$ \\
	\hline 
	& $\overline{X}_1$ & $X$ & $X$ & $X$ & $X$ & $I$ & $I$ & $I$ & $I$ & $I$ & $I$ & $I$ & $I$ \\
	& $\overline{Z}_1$ & $Z$ & $I$ & $I$ & $I$ & $Z$ & $I$ & $I$ & $I$ & $I$ & $I$ & $I$ & $I$ \\
	\hline 
	& $\overline{X}_2$ & $X$ & $X$ & $I$ & $I$ & $X$ & $X$ & $I$ & $I$ & $I$ & $I$ & $I$ & $I$ \\
	& $\overline{Z}_2$ & $Z$ & $I$ & $Z$ & $I$ & $I$ & $I$ & $I$ & $I$ & $I$ & $I$ & $I$ & $I$ \\
	\hline 
	& $\overline{X}_3$ & $X$ & $I$ & $X$ & $I$ & $X$ & $I$ & $X$ & $I$ & $I$ & $I$ & $I$ & $I$ \\
	& $\overline{Z}_3$ & $Z$ & $Z$ & $I$ & $I$ & $I$ & $I$ & $I$ & $I$ & $I$ & $I$ & $I$ & $I$ \\
	\hline 
	& $\overline{X}_4$ & $I$ & $I$ & $I$ & $I$ & $I$ & $I$ & $I$ & $I$ & $X$ & $I$ & $X$ & $I$ \\
	& $\overline{Z}_4$ & $I$ & $I$ & $I$ & $I$ & $I$ & $I$ & $I$ & $I$ & $Z$ & $Z$ & $I$ & $I$ \\
	\hline 
	& $\overline{X}_5$ & $I$ & $I$ & $I$ & $I$ & $I$ & $I$ & $I$ & $I$ & $X$ & $X$ & $I$ & $I$ \\
	& $\overline{Z}_5$ & $I$ & $I$ & $I$ & $I$ & $I$ & $I$ & $I$ & $I$ & $Z$ & $I$ & $Z$ & $I$ \\
	\hline
    \end{tabular}
 \caption{Stabilisers of the $\SICC$ and $\nkd{4}{2}{2}$ code.}
 \label{tab:teleport-Hadamard}
\end{table}

\subsection{Interaction with the $\nkd{4}{2}{2}$ code}

Due to the inherent difficulty in directly implementing logical Hadamard gates on the encoded qubits of the $\SICC$ code, it is natural to explore an alternative approach.
One potential solution involves teleporting quantum states from one code to another, where the implementation of logical Hadamard gates is easier.
The resulting state can then be teleported back to the original $\SICC$ code block, completing the process. 

\begin{figure}
  \centering
  \resizebox{0.48\textwidth}{!}{\begin{tikzpicture}[y = 1cm, x = 1cm, inner sep = 0pt, outer sep = 0pt, every path/.style={draw=black, line cap=round, line join=round, line width=0.05cm}]

\begin{scope}[shift={(0,0)}]
	\path[fill = tikzwhite, fill opacity = 0.8] (2,2) -- (6,2) -- (6,6) -- (2,6) -- cycle; 
	\path[fill = tikzwhite, fill opacity = 0.8] (0,0) -- (0,4) -- (2,6) -- (2,2) -- cycle; 
	\path[fill = tikzwhite, fill opacity = 0.8] (0,0) -- (4,0) -- (6,2) -- (2,2) -- cycle; 
	\path[fill = tikzwhite, fill opacity = 0.8] (4,0) -- (4,4) -- (6,6) -- (6,2) -- cycle; 
	\path[fill = tikzwhite, fill opacity = 0.8] (0,4) -- (4,4) -- (6,6) -- (2,6) -- cycle; 
	\path[fill = tikzwhite] (2,6) circle (0.5cm);
	\node at (2,6) {\fontsize{1cm}{1.2cm}$6$}; 
	\path[fill = tikzwhite] (2,2) circle (0.5cm);
	\node at (2,2) {\fontsize{1cm}{1.2cm}$7$}; 
	\path[fill = tikzwhite, fill opacity = 0.8] (0,0) -- (4,0) -- (4,4) -- (0,4) -- cycle; 
	\path[fill = tikzwhite] (0,4) circle (0.5cm);
	\node at (0,4) {\fontsize{1cm}{1.2cm}$2$}; 
	\path[fill = tikzwhite] (4,4) circle (0.5cm);
	\node at (4,4) {\fontsize{1cm}{1.2cm}$0$}; 
	\path[fill = tikzwhite] (6,6) circle (0.5cm);
	\node at (6,6) {\fontsize{1cm}{1.2cm}$4$}; 
	\path[fill = tikzwhite] (6,2) circle (0.5cm);
	\node at (6,2) {\fontsize{1cm}{1.2cm}$5$}; 
	\path[fill = tikzwhite] (4,0) circle (0.5cm);
	\node at (4,0) {\fontsize{1cm}{1.2cm}$1$}; 
	\path[fill = tikzwhite] (0,0) circle (0.5cm);
	\node at (0,0) {\fontsize{1cm}{1.2cm}$3$}; 
	\path[fill = tikzwhite, fill opacity = 0.8] (0,8) -- (4,8) -- (6,10) -- (2,10) -- cycle; 
	\path[fill = tikzwhite] (0,8) circle (0.5cm);
	\node at (0,8) {\fontsize{1cm}{1.2cm}$9$}; 
	\path[fill = tikzwhite] (4,8) circle (0.5cm);
	\node at (4,8) {\fontsize{1cm}{1.2cm}$8$}; 
	\path[fill = tikzwhite] (6,10) circle (0.5cm);
	\node at (6,10) {\fontsize{1cm}{1.2cm}$10$}; 
	\path[fill = tikzwhite] (2,10) circle (0.5cm);
	\node at (2,10) {\fontsize{1cm}{1.2cm}$11$}; 
\end{scope}

\begin{scope}[shift={(12, 0)}]
	\path[fill = tikzwhite, fill opacity = 0.8] (2,2) -- (6,2) -- (6,6) -- (2,6) -- cycle; 
	\path[fill = tikzwhite, fill opacity = 0.8] (0,0) -- (0,4) -- (2,6) -- (2,2) -- cycle; 
	\path[fill = tikzwhite, fill opacity = 0.8] (0,0) -- (4,0) -- (6,2) -- (2,2) -- cycle; 
	\path[fill = tikzwhite, fill opacity = 0.8] (4,0) -- (4,4) -- (6,6) -- (6,2) -- cycle; 
	\path[fill = tikzwhite, fill opacity = 0.8] (0,4) -- (4,4) -- (6,6) -- (2,6) -- cycle; 
	
	\path[fill = tikzwhite, fill opacity = 0.8] (0,0) -- (4,0) -- (4,4) -- (0,4) -- cycle; 
	
	\path[fill = tikzblack] (0,4) circle (0.2cm);
	\path[line width=0.075cm] (0,4) --++ (0,4);
	\path[fill = tikzblack] (4,4) circle (0.2cm);
	\path[line width=0.075cm] (4,4) --++ (0,4);
	\path[fill = tikzblack] (6,6) circle (0.2cm);
	\path[line width=0.075cm] (6,6) --++ (0,4);
	\path[fill = tikzblack] (2,6) circle (0.2cm);
	\path[line width=0.075cm] (2,6) --++ (0,4);

	\path[fill = tikzwhite, line width=0.075cm] (6,10) circle (0.5cm);
	\path[line width=0.075cm] (6,9.5) --++ (0,1);
	\path[line width=0.075cm] (5.5,10) --++ (1,0);
	\path[fill = tikzwhite, line width=0.075cm] (2,10) circle (0.5cm);
	\path[line width=0.075cm] (2,9.5) --++ (0,1);
	\path[line width=0.075cm] (1.5,10) --++ (1,0);

	\path[fill = tikzwhite, fill opacity = 0.8] (0,8) -- (4,8) -- (6,10) -- (2,10) -- cycle; 
	\path[fill = tikzwhite, line width=0.075cm] (0,8) circle (0.5cm);
	\path[line width=0.075cm] (0,7.5) --++ (0,1);
	\path[line width=0.075cm] (-0.5,8) --++ (1,0);
	\path[fill = tikzwhite, line width=0.075cm] (4,8) circle (0.5cm);
	\path[line width=0.075cm] (4,7.5) --++ (0,1);
	\path[line width=0.075cm] (3.5,8) --++ (1,0);
\end{scope}

\begin{scope}[shift={(0, -12)}]
	\path[fill = tikzwhite, fill opacity = 0.8] (2,2) -- (6,2) -- (6,6) -- (2,6) -- cycle; 
	\path[fill = tikzwhite, fill opacity = 0.8] (0,0) -- (0,4) -- (2,6) -- (2,2) -- cycle; 
	\path[fill = tikzwhite, fill opacity = 0.8] (0,0) -- (4,0) -- (6,2) -- (2,2) -- cycle; 
	\path[fill = tikzyellow, fill opacity = 0.8] (4,0) -- (4,4) -- (6,6) -- (6,2) -- cycle; 
	\path[fill = tikzwhite, fill opacity = 0.8] (0,4) -- (4,4) -- (6,6) -- (2,6) -- cycle; 
	\path[fill = tikzwhite, fill opacity = 0.8] (0,0) -- (4,0) -- (4,4) -- (0,4) -- cycle; 
	
	\node at (5,3) {\logicallabel{$\overline{X}_2$}};

	\path[fill = tikzwhite, fill opacity = 0.8] (0,8) -- (4,8) -- (6,10) -- (2,10) -- cycle; 
	
	\path[draw = tikzmagenta, draw opacity = 0.8, line width = 1.2cm] (0,8) -- (4,8);
	\node at (2,8) {\logicallabel{$\overline{Z}_4$}};
	
	\begin{scope}[shift={(9, 7)}]
	\path[-{To[length=5mm]}] (-2, 0) -- (2, 0);
	\begin{scope}[shift={(-1, -3.125)}, scale=0.25]
		\path[fill = tikzwhite, fill opacity = 0.8] (2,2) -- (6,2) -- (6,6) -- (2,6) -- cycle; 
		\path[fill = tikzwhite, fill opacity = 0.8] (0,0) -- (0,4) -- (2,6) -- (2,2) -- cycle; 
		\path[fill = tikzwhite, fill opacity = 0.8] (0,0) -- (4,0) -- (6,2) -- (2,2) -- cycle; 
		\path[fill = tikzwhite, fill opacity = 0.8] (4,0) -- (4,4) -- (6,6) -- (6,2) -- cycle; 
		\path[fill = tikzwhite, fill opacity = 0.8] (0,4) -- (4,4) -- (6,6) -- (2,6) -- cycle; 
		\path[fill = tikzwhite, fill opacity = 0.8] (0,0) -- (4,0) -- (4,4) -- (0,4) -- cycle; 
		\path[fill = tikzwhite, line width=0.0375cm] (6,10) circle (0.5cm);
		\path[line width=0.0375cm] (6,9.5) --++ (0,1);
		\path[line width=0.0375cm] (5.5,10) --++ (1,0);
		\path[fill = tikzwhite, line width=0.0375cm] (2,10) circle (0.5cm);
		\path[line width=0.0375cm] (2,9.5) --++ (0,1);
		\path[line width=0.0375cm] (1.5,10) --++ (1,0);

		\path[fill = tikzwhite, fill opacity = 0.8] (0,8) -- (4,8) -- (6,10) -- (2,10) -- cycle; 
		\path[fill = tikzwhite, line width=0.0375cm] (0,8) circle (0.5cm);
		\path[line width=0.0375cm] (0,7.5) --++ (0,1);
		\path[line width=0.0375cm] (-0.5,8) --++ (1,0);
		\path[fill = tikzwhite, line width=0.0375cm] (4,8) circle (0.5cm);
		\path[line width=0.0375cm] (4,7.5) --++ (0,1);
		\path[line width=0.0375cm] (3.5,8) --++ (1,0);

		\path[fill = tikzblack] (0,4) circle (0.2cm);
		\path[line width=0.0375cm] (0,4) --++ (0,4);
		\path[fill = tikzblack] (4,4) circle (0.2cm);
		\path[line width=0.0375cm] (4,4) --++ (0,4);
		\path[fill = tikzblack] (6,6) circle (0.2cm);
		\path[line width=0.0375cm] (6,6) --++ (0,4);
		\path[fill = tikzblack] (2,6) circle (0.2cm);
		\path[line width=0.0375cm] (2,6) --++ (0,4);
	\end{scope}
	\end{scope}

	\begin{scope}[shift={(12, 0)}]
	\path[fill = tikzwhite, fill opacity = 0.8] (2,2) -- (6,2) -- (6,6) -- (2,6) -- cycle; 
	\path[fill = tikzwhite, fill opacity = 0.8] (0,0) -- (0,4) -- (2,6) -- (2,2) -- cycle; 
	\path[fill = tikzwhite, fill opacity = 0.8] (0,0) -- (4,0) -- (6,2) -- (2,2) -- cycle; 
	\path[fill = tikzwhite, fill opacity = 0.8] (0,4) -- (4,4) -- (6,6) -- (2,6) -- cycle; 
	\path[fill = tikzwhite, fill opacity = 0.8] (4,0) -- (4,4) -- (6,6) -- (6,2) -- cycle; 
	\path[draw=none, fill = tikzred, fill opacity = 0.8] (4, 0) -- (4,8) -- (6,10) -- (6,2) -- cycle;
	\path[fill = tikzwhite, fill opacity = 0.8] (0,0) -- (4,0) -- (4,4) -- (0,4) -- cycle; 
	\path[fill = tikzwhite, fill opacity = 0.8] (0,8) -- (4,8) -- (6,10) -- (2,10) -- cycle; 
	\path[draw=none, fill = tikzred, fill opacity = 0.8] (0,4) -- (4,4) -- (4,8) -- (0,8) -- cycle;
	\node at (5,3) {\logicallabel{$\overline{X}_2 \overline{X}_4$}};
	\node at (2,7) {\logicallabel{$\overline{Z}_2 \overline{Z}_4$}};
	\end{scope}

\end{scope}

\end{tikzpicture}}
  \caption{Transversal logical \cnot{} between a face of the $\SICC$ and $\nkd{4}{2}{2}$ codes.
    $\overline{Z}$ operators contained in the specified face, and $\overline{X}$ operators which share an edge with that face are affected. 
  }
  \label{fig:interaction}
\end{figure}
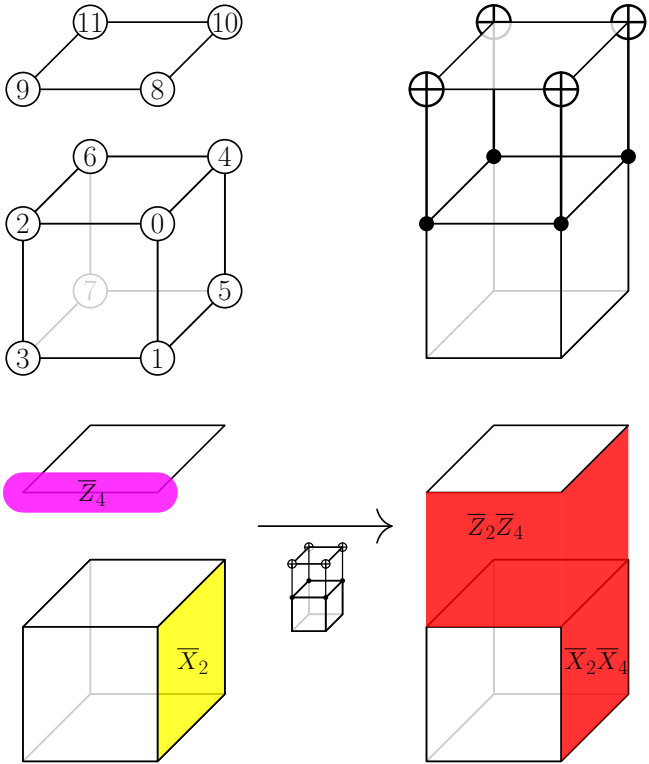

To serve as the target for the teleportation process, an advantageous choice is a $\nkd{4}{2}{2}$ error-detecting code, with its corresponding stabilizers and logical operators listed in \cref{tab:teleport-Hadamard}.
The selection of the $\nkd{4}{2}{2}$ code is motivated by two crucial factors.
Firstly, the logical $\cnot$ between two encoded qubits of the $\SICC$ code and the two encoded qubits of the $\nkd{4}{2}{2}$ is transversal (see \cref{fig:interaction}).
This transversal gate greatly facilitates teleportation between these two codes.
Secondly, the $\nkd{4}{2}{2}$ code offers the advantage of easily preparing eigenstates of logical Pauli operators \cite{Knill2005}.

The presence of two logical qubits in the $\nkd{4}{2}{2}$ block allows the implementation of two simultaneous logical Hadamard gates through teleportation, we describe the circuit for this operation in the following subsection, before concluding with the implementation of a single logical Hadamard.

\subsection{Double logical Hadamard}
\begin{figure}
\includegraphics[width=0.48\textwidth]{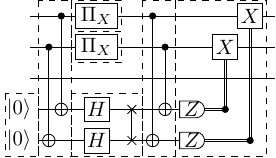}
\caption{\label{fig:logical-double-hadamard-full}Teleportation-based logical circuit for performing two Hadamard gates simultaneously.
Operations in dashed boxes can be implemented using fault-tolerant subcircuits.}
\end{figure}

\begin{figure*}[!htpb]
\includegraphics[width=0.98\textwidth]{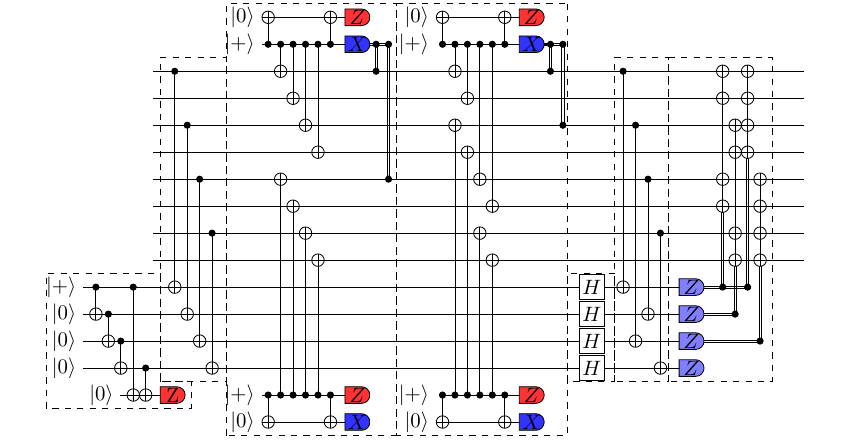}
\caption{\label{fig:physical-double-hadamard-full}Physical circuit corresponding to \cref{fig:logical-double-hadamard-full}, with fault-tolerant subcircuits in dashed boxes. }
\end{figure*}

Here, we consider a logical Hadamard gate acting on logical qubits 1 and 2 of the $\SICC$ code simultaneously.
Note that, by symmetry, any other pair of logical qubits could be chosen, and the resulting circuit would be identical up to a permutation of physical qubits.
At the logical level, this consists of preparing the $\ket{\overline{0}\overline{0}}$ state of the $\nkd{4}{2}{2}$ code, performing the transversal \cnot{}, non-destructively projecting logical qubits 1 and 2 to the state $\ket{\overline{+}\overline{+}}$, carrying out the logical Hadamard on the $\nkd{4}{2}{2}$ block, and teleporting the resulting states back to the $\SICC$ block, using destructive logical measurement on the unneeded $\nkd{4}{2}{2}$ block (see \cref{fig:logical-double-hadamard-full}). 
Each of these operations can be implemented at the physical level using transversal gates and flag circuits, without requiring a QEC gadget (see \cref{fig:physical-double-hadamard-full}).

\subsection{Single logical Hadamard}
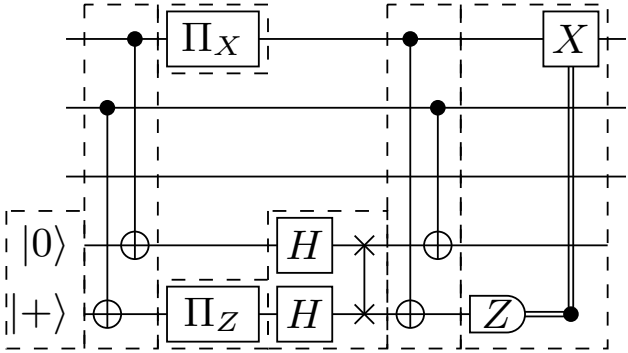
\begin{figure}
\centering
\resizebox{0.48\textwidth}{!}{\begin{tikzpicture}[scale=1,x=1pt,y=1pt]
\pgfmathsetmacro{\xmin}{-4}
\pgfmathsetmacro{\xmax}{118}
\filldraw[color=white] (\xmin,  -7.5) rectangle (114, 67.5);
\draw[color=black] (\xmin, 60) -- (\xmax,60);
\draw[color=black] (\xmin, 45) -- (\xmax,45);
\draw[color=black] (\xmin, 30) -- (\xmax,30);
\draw[color=black] (0, 15) -- (114,15);
\draw[color=black] (0, 15) node[left] {$\ket{0}$};
\draw[color=black] (0, 0) -- (90,0);
\draw[color=black] (90,-0.5) -- (106,-0.5);
\draw[color=black] (90,0.5) -- (106,0.5);
\draw[color=black] (0, 0) node[left] {$\ket{+}$};
\draw (5,45) -- (5,0);
\begin{scope}
\draw[fill=white] (5, 0) circle(3pt);
\clip (5, 0) circle(3pt);
\draw (2, 0) -- (8, 0);
\draw (5, -3) -- (5, 3);
\end{scope}
\filldraw (5, 45) circle(1.5pt);
\draw (11,60) -- (11,15);
\begin{scope}
\draw[fill=white] (11, 15) circle(3pt);
\clip (11, 15) circle(3pt);
\draw (8, 15) -- (14, 15);
\draw (11, 12) -- (11, 18);
\end{scope}
\filldraw (11, 60) circle(1.5pt);
\begin{scope}
\draw[fill=white] (28, 60) +(-45:14.142136pt and 8.485281pt) -- +(45:14.142136pt and 8.485281pt) -- +(135:14.142136pt and 8.485281pt) -- +(225:14.142136pt and 8.485281pt) -- cycle;
\clip (28, 60) +(-45:14.142136pt and 8.485281pt) -- +(45:14.142136pt and 8.485281pt) -- +(135:14.142136pt and 8.485281pt) -- +(225:14.142136pt and 8.485281pt) -- cycle;
\draw (28, 60) node {$\Pi_X$};
\end{scope}
\begin{scope}
\draw[fill=white] (28, -0) +(-45:14.142136pt and 8.485281pt) -- +(45:14.142136pt and 8.485281pt) -- +(135:14.142136pt and 8.485281pt) -- +(225:14.142136pt and 8.485281pt) -- cycle;
\clip (28, -0) +(-45:14.142136pt and 8.485281pt) -- +(45:14.142136pt and 8.485281pt) -- +(135:14.142136pt and 8.485281pt) -- +(225:14.142136pt and 8.485281pt) -- cycle;
\draw (28, -0) node {$\Pi_Z$};
\end{scope}
\begin{scope}
\draw[fill=white] (48, 15) +(-45:8.485281pt and 8.485281pt) -- +(45:8.485281pt and 8.485281pt) -- +(135:8.485281pt and 8.485281pt) -- +(225:8.485281pt and 8.485281pt) -- cycle;
\clip (48, 15) +(-45:8.485281pt and 8.485281pt) -- +(45:8.485281pt and 8.485281pt) -- +(135:8.485281pt and 8.485281pt) -- +(225:8.485281pt and 8.485281pt) -- cycle;
\draw (48, 15) node {$H$};
\end{scope}
\begin{scope}
\draw[fill=white] (48, -0) +(-45:8.485281pt and 8.485281pt) -- +(45:8.485281pt and 8.485281pt) -- +(135:8.485281pt and 8.485281pt) -- +(225:8.485281pt and 8.485281pt) -- cycle;
\clip (48, -0) +(-45:8.485281pt and 8.485281pt) -- +(45:8.485281pt and 8.485281pt) -- +(135:8.485281pt and 8.485281pt) -- +(225:8.485281pt and 8.485281pt) -- cycle;
\draw (48, -0) node {$H$};
\end{scope}
\draw (61,15) -- (61,0);
\begin{scope}
\draw (58.878680, 12.878680) -- (63.121320, 17.121320);
\draw (58.878680, 17.121320) -- (63.121320, 12.878680);
\end{scope}
\begin{scope}
\draw (58.878680, -2.121320) -- (63.121320, 2.121320);
\draw (58.878680, 2.121320) -- (63.121320, -2.121320);
\end{scope}
\draw (71,60) -- (71,0);
\begin{scope}
\draw[fill=white] (71, 0) circle(3pt);
\clip (71, 0) circle(3pt);
\draw (68, 0) -- (74, 0);
\draw (71, -3) -- (71, 3);
\end{scope}
\filldraw (71, 60) circle(1.5pt);
\draw (77,45) -- (77,15);
\begin{scope}
\draw[fill=white] (77, 15) circle(3pt);
\clip (77, 15) circle(3pt);
\draw (74, 15) -- (80, 15);
\draw (77, 12) -- (77, 18);
\end{scope}
\filldraw (77, 45) circle(1.5pt);
\draw[fill=white] (84, -4) -- (92,-4) arc (-90:90:4pt) -- (84,4) -- cycle;
\draw (90, 0) node {$Z$};
\draw (105.5,60) -- (105.5,0);
\draw (106.5,60) -- (106.5,0);
\begin{scope}
\draw[fill=white] (106, 60) +(-45:8.485281pt and 8.485281pt) -- +(45:8.485281pt and 8.485281pt) -- +(135:8.485281pt and 8.485281pt) -- +(225:8.485281pt and 8.485281pt) -- cycle;
\clip (106, 60) +(-45:8.485281pt and 8.485281pt) -- +(45:8.485281pt and 8.485281pt) -- +(135:8.485281pt and 8.485281pt) -- +(225:8.485281pt and 8.485281pt) -- cycle;
\draw (106, 60) node {$X$};
\end{scope}
\filldraw (106, 0) circle(1.5pt);

\draw[draw opacity=1,fill opacity=0.2,color=black,,dashed] (-17,22.5) rectangle (0,-7.5);

\draw[draw opacity=1,fill opacity=0.2,color=black,,dashed] (0,67.5) rectangle (16,-7.5);

\draw[draw opacity=1,fill opacity=0.2,color=black,,dashed] (16,67.5) rectangle (40,52.5);

\draw[draw opacity=1,fill opacity=0.2,color=black,,dashed] (16,7.5) rectangle (40,-7.5);

\draw[draw opacity=1,fill opacity=0.2,color=black,,dashed] (40,22.5) rectangle (66,-7.5);

\draw[draw opacity=1,fill opacity=0.2,color=black,,dashed] (66,67.5) rectangle (82,-7.5);

\draw[draw opacity=1,fill opacity=0.2,color=black,,dashed] (82,-7.5) rectangle (114,67.5);

\end{tikzpicture}}
\caption{Teleportation-based Hadamard for the $\SICC$ code represented at the logical level. 
The top three wires represent the three logical qubits of the $\SICC$ code, with the bottom two representing the logical qubits of the $\nkd{4}{2}{2}$ code.
Sets of operations surrounded with dashed boxes can be carried out using fault-tolerant sub-circuits.}
\label{fig:logical-hadamard-full}
\end{figure}

To implement a logical Hadamard gate on a single qubit, we change the initial state on the $\nkd{4}{2}{2}$ code to $\ket{\overline{0}\overline{+}}$, and project only one logical qubit of the $\SICC$ code into the $\ket{\overline{+}}$ state. 
Otherwise, the circuits implementing this gate are quite similar at both the logical and physical levels, see \cref{fig:logical-hadamard-full} and \cref{fig:physical-hadamard-full}, respectively.

\begin{figure*}
\resizebox{0.98\textwidth}{!}{\input{physical_hadamard_full_touchup.tikz}}
\caption{Teleportation-based Hadamard for the $\SICC$ code represented at the physical level. 
The eight wires with no fixed input represent the physical qubits of the $\SICC$ code, with the four immediately beneath them representing the physical qubits of the $\nkd{4}{2}{2}$ code.
Sets of operations surrounded with dashed boxes correspond to operations in \cref{fig:logical-hadamard-full}.
Measurements highlighted in red detect propagating gate errors; if they return a non-trivial result, the computation is restarted.
Sets of measurements highlighted in shades of blue must return results of even parity for the computation to continue.}
\label{fig:physical-hadamard-full}
\end{figure*}

\end{document}